# Nondestructive testing of high strength conductors for high field pulsed magnets

Jun Lu, Todd Adkins, Iain Dixon (senior IEEE member), Doan Nguyen and Ke Han

*Abstract*— **High field pulsed magnets at the NHMFL use high strength conductor wires up to 90% of their ultimate tensile strength. Therefore it is very important to ensure that the wires are free of flaws. It is known that in the conductors' cold drawing process, internal 'chevron' crack could occur due to unsuitable drawing die schedule or inadequate lubrication. These internal cracks occurs infrequently along the wire, so tensile tests of short samples cut from the ends of a long length conductor often miss the problem. In addition, small inclusions on the wire surface can compromise wires fatigue properties.**

**In this paper, we present results of our non-destructive testing (NDT) inspection of Glidcop AL60 wires using eddy current testing (ECT), ultrasonic testing (UT) and x-ray radiography (2D and 3D). Chevron cracks were found in some AL60 conductors by all three NDT techniques. Surface inclusions were found by ECT. We have developed a long length ECT wire inspection capability.**

## I. Introduction

High field pulsed magnet requires high strength conductors to withstand large electromagnetic force generated by combination of high electrical current and high magnetic field. Several high strength conductors have been developed specifically for this application [1]-[3]. While development efforts are being vigorously made to improve conductor strength, wires of existing materials are used in designs with stress up to 90% of the ultimate tensile strength [4]. At such high stress, small flaws on the conductor can result in catastrophic magnet failure. Evidently the quality assurance tests that make sure the wires are flaw-free is absolutely crucial.

Previously at the NHMFL, short samples were cut from front and back ends of a long wire. Mechanical and electrical characterization were performed on the short samples. The wires were also visually inspected for surface flaws. Recently a Glidcop AL60 wire broke in the drawing process indicating severe internal cracks prior to the breakage. It was suspected that internal cracks also exist in some finished wires. Therefore it is prudent to develop a technology to inspect all the wires for such internal cracks.

It is well-known that internal cracks occur sporadically in wire cold drawing process due to unsuitable die schedule and/or poor lubrication [5]-[8]. These cracks, also called central bursts or 'chevron' cracks, occurs infrequently along the wire, tensile tests of short samples often miss the problem.

In addition, small surface flaws and metallic inclusions can compromise the fatigue properties of the wire and likely reduce the number of pulses (shots) available in a magnet's lifetime. But they are usually very difficult to discern visually. Therefore it is also important to develop a method for better surface inspection.

The goal of this research is to identify suitable nondestructive testing (NDT) methods [9] for inspection of both internal and surface flaws, which are compatible in scales and costs with pulsed magnet fabrication process [10]. In this paper, we present our results on inspection of a short Glidcop AL60 (AL60) sample using eddy current testing (ECT), ultrasonic testing (UT) and x-ray radiography and computed tomography (CT). Results from ECT inspections of long length wires will also be presented.

## II. Experimental

AL60 precursors made by North American Höganäs High Alloys LLC. were in the form of hot-extruded 13.7 or 20.0 mm diameter rod. The finished rectangular shaped wire of various cross-sections were drawn by Sam Dong Inc.

An Olympus NORTEC 600 eddy current flaw detector was used for ECT inspection using several probes. A 0.44" diameter 500 Hz- 40 kHz spot probe, a 0.125" diameter 50 kHz – 199 KHz surface probes, and a 0.82" inner diameter 100 kHz – 500 KHz encircling probe were used for different types of flaws. Plastic fixtures were made to hold and guide the probes on different size of wires during inspection.

UT inspection was performed by an Olympus EPOCH 650 ultrasonic flaw detector with a DHC711 5 MHz dual-element longitudinal wave transducer coupled to samples by glycerin gel.

X-ray inspection was performed by Delphi Precision Imaging Inc. where a North Star Imaging X5000 system equipped with a 225 kV micro-focus x-ray tube was used. The estimated maximum lateral resolution in this work is about 20 μm. Both 2D radiography scans and 3D computed tomography (CT) scans were performed.

## III. Results and Discussions

### A. Basic considerations of NDT techniques

#### 1. ECT

When an ac excitation coil is in the vicinity of a wire, the impedance of the coil is influenced by the induced eddy current

The NHMFL is supported by NSF through NSF-DMR-1157490 and 1644779, and the State of Florida. Jun Lu, Todd Adkins, Iain Dixon and Ke Han are with the National High Magnetic Field Laboratory, Tallahassee, FL 32310, USA.

Doan Nguyen is with National High Magnetic Field Laboratory, Pulsed Field Facility, Los Alamos National Laboratory, NM 87545, USA (Corresponding author: Jun Lu, junlu@magnet.fsu.edu).

in the wire. A flaw interrupts eddy current flow and changes the impedance, measurement of which is then used for flaw detection. Due to the well-known skin effect, eddy current decays from wire surface exponentially. The penetration depth $\delta$ describe this exponential decay can be written as a function of ac frequency $f$,

$$\delta = \sqrt{\frac{\rho}{\mu \pi f}}$$

where $\rho$ is the electrical resistivity, $\mu$ the magnetic permeability. For AL60, with its electrical conductivity of 96% of international annealed copper standard (IACS), it can be calculated that penetration depths are 2.95 and 0.30 mm for 500 Hz and 50 kHz respectively.

An experiment was designed to confirm the skin effect in AL60 wire. A hole of 2 mm diameter and 1.2 mm below the front surface was drilled from the backside of a 5.2 mm thick AL60 wire. The 0.44" diameter 500 Hz – 40 kHz spot probe was used. Fig. 2 shows the ECT signal of the hole as a function of operating frequency. Evidently, the signal decreases with increasing frequency which corresponds to decreasing eddy current penetration depth. It is significantly smaller when the penetration frequency is smaller than the depth of the hole of 1.2 mm.

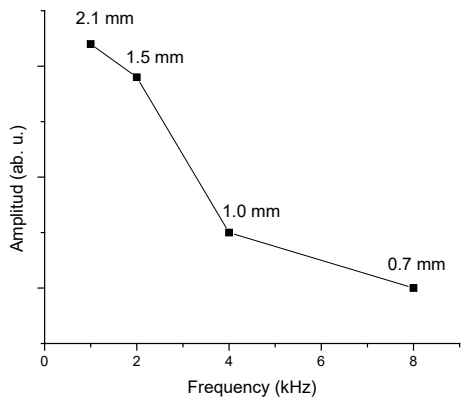

Fig. 1 ECT flaw signal amplitude versus frequency for a hole of 2 mm diameter and 1.2 mm below the surface of an AL60 wire. The eddy current penetration depths in mm are labelled for each frequency. The solid line is a guide to the eye.

Most commercial ECT probes operates above or at 100 Hz. At this frequency, penetration depth for copper is about 6.6 mm. This seems to be the penetration limit of ECT for our application. On the other hand, frequency dependence of eddy current penetration can be used to estimate the depth of a flaw. As indicated in Fig. 1, it can be estimated that the flaw is deeper than 1 mm. Our experiments also showed that lateral spatial resolution of ECT is in the order of millimeters, and it is difficult to estimate the size and shape of a flaw. In spite of these limitations, it seems to be possible to develop a criterion based on the amplitude of the flaw signal and use it to accept or reject wires.

For inspection of large volume of long wires, ECT seems to be an efficient and low cost method. It also has the advantage of sensitive to surface cracks and inclusions especially if inclusions consist of ferromagnetic steel which is quite common.

*2. UT*

UT generates ultrasonic waves and uses the reflection of ultrasonic waves to detect a flaw. High frequency ultrasound provide better spatial resolution at a cost of reduced penetration depth due to the sound wave attenuation in the material. The ultrasonic wave attenuation in AL60 was measured by an echo amplitude vs. time (A-scan) of an 8.4 mm thick wire which showed multiple back wall reflections. As shown in Fig. 2, the back wall reflection amplitudes decay with distance. This decay is consistent with an exponential decay function with a characteristic length of about 15 mm (the dashed line). It shows that UT can easily penetrate 20 mm of AL60 material with the ultrasonic power used in that experiment. This is adequate for inspection of our AL60 wires and precursor rods.

The main disadvantage of UT is its lack of surface detection ability. In addition, the need to use couplant between the transducer and the wire makes long length inspection impractical. An immersion UT [9] technique might be able to overcome this disadvantage, which requires further development.

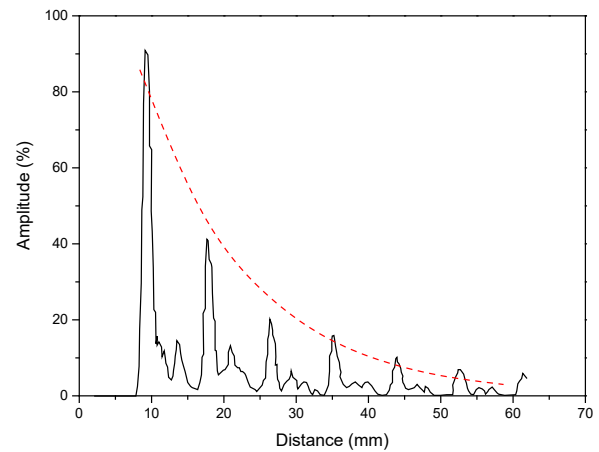

Fig. 2. UT echo amplitude versus distance (A-scan) of an 8.4 mm thick AL60 wire. The peaks are multiple back well reflections. The distance between peaks is 8.4 mm. The dashed line is an exponential function with a characteristic length of 15 mm.

*3. X-ray*

X-ray radiography (2D) and computed tomography (CT) have the ability to determine flaw's shape and size with high spatial resolution. Its lateral resolution is controlled by the spot size of the x-ray source. For this work, the spot size was chosen so that 13 - 28 μm resolution can be achieved. The x-ray penetration depth depends on the voltage and spot size of the x-ray source, the sensitivity of the detector, as well as the absorption coefficient of material being inspected. In case of AL60, at 225 kV the x-ray source was able to penetrate and image a 12.3 mm thick sample.

The investment for an x-ray system is high; the training required for operating seems to be long; its use for inspection of long wires seems to be difficult. In addition, the health hazardous of intense radiation is a concern. Therefore its use is limited to some specialized facilities. In this work, x-ray was used to accurately measure the locations and sizes of internal flaws, in order to provide a reference for developments of ECT and UT methods.

### B. Surface cracks during coil winding

Surface cracks were observed on one occasion when an uninsulated AL60 wire was being wound to a coil. Apparently, the cracks were created by the bending stress. For insulated wires, it is not possible to visually observe cracks under the insulation. So NDT is needed. This was when our first ECT inspection was developed and applied to pulsed magnet wire.

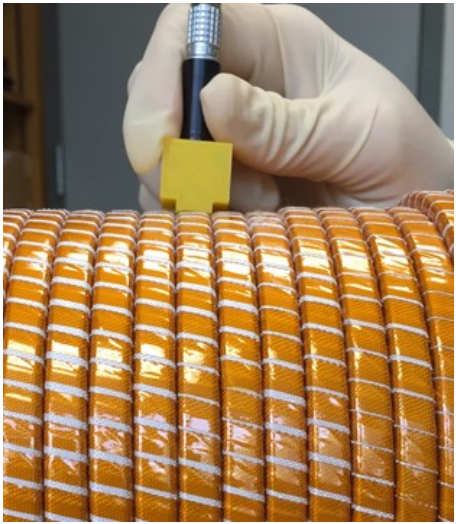

Fig. 3. ECT inspection of surface cracks during an AL60 coil winding. An Olympus 0.125" diameter 50 – 199 kHz surface probe was used.

Fig. 3 depicts the ECT inspection during a coil winding. High frequency probe up to 199 kHz was used for high surface sensitivity. The corresponding penetration depth is 0.15 mm. The outer surface of entire wire was successfully inspected as the wire was wound into a coil. No cracks were found.

### C. Chevron cracks

AL60 wire is an excellent conductor with a combination of high strength and high conductivity. But it is challenging to draw without breakage which is often associated with chevron cracks created prior to the breakage. In one case during our coil winding, the wire snapped. When a sample adjacent to the breakage was inspected with ECT, a large ECT flaw signal was discovered.

In order to systematically study the chevron cracks, a wire of 40 cm long 8.4 x 12.3 mm$^2$ cross-section were sent to Delphi Precision Imaging Inc. for x-ray inspections. This wire has a possible chevron crack as indicated by a very shallow dent on the surface. Indeed, x-ray 2D and CT scans revealed a large chevron cracks as shown in Fig. 4. Two additional chevron cracks with no surface indications were also found, as labeled A and B in Fig. 5. This sample with known chevron cracks identified by x-ray were subsequently used as a reference sample for ECT and UT studies.

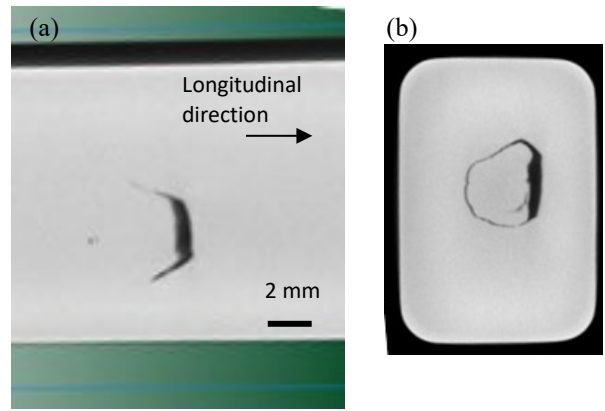

Fig. 4. X-ray computed tomography (CT) results of a large chevron crack.in an 8.4 x 12.3 mm$^2$ AL60 wire. (a) longitudinal slice of the CT data. (b) transverse cross-section slice of the CT data.

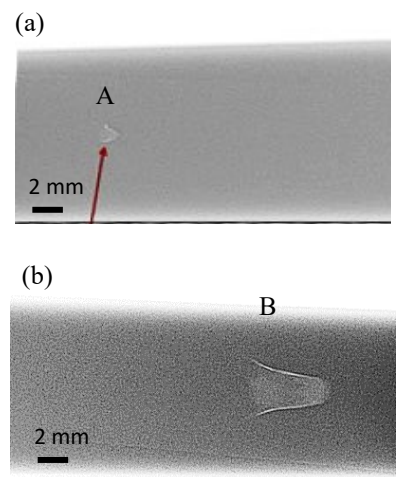

Fig. 5 2D x-ray radiographs of the 12.3 x 8.4 mm$^2$ AL60 wire (a) a small chevron crack A and (b) a large chevron crack B.

ECT was performed on both flaw A and B at 500 Hz. The results are shown in Fig. 6. There was no significant ECT signal from the small flaw A (Fig. 6(a)). This is likely due to its smaller size. Because the minimum depth of this flaw is greater than the penetration depth of 2.95 mm at 500 Hz. In contrast, the minimum depth of the much larger flaw B is less than 2.95 mm. Therefore a significant ECT signal was observed as shown in Fig. 6(b). These results are qualitatively in agreement with x-ray images.

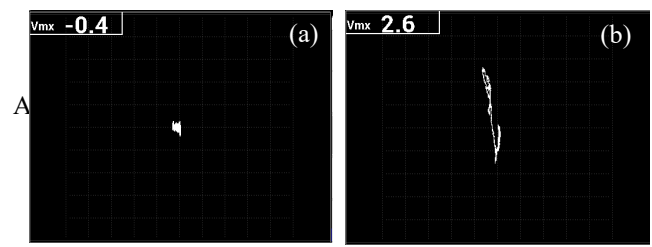

Fig. 6. ECT signal of (a) flaw A and (b) flaw B at 500 Hz.

UT A-scan was also performed on flaws A and B of the same sample. A-scan traces from regions with flaw A and B are shown in Fig. 7 together with a trace on a no-flaw region as a reference. The trace of no-flaw region shows only reflections from back wall. In comparison, the trace of flaw A region has significantly reduced back wall reflections, and a broad peak appears at about half the time of the first back wall reflection, indicating a reflection from flaw A, which is near the center of the wire in depth. In the trace of flaw B, the back wall reflection is very small. A peak on the left of the flaw A reflection indicates that the top of the flaw is further close to the surface.

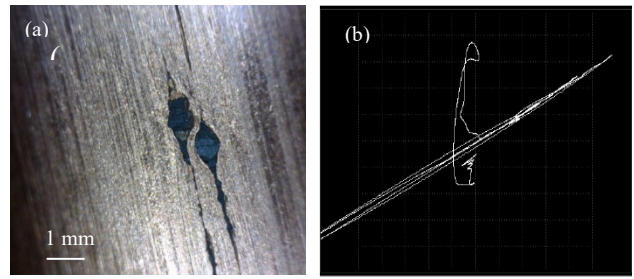

Fig. 8. Inspection of surface inclusions on an AL60 wire (a) a picture of two inclusions and (b) the corresponding ECT signal.

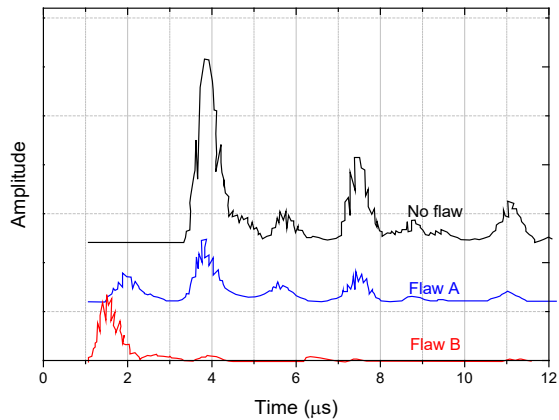

Fig. 7. UT A-scan of the two flaws shown in Fig. 5.

Data in Fig. 7 demonstrates that an internal flaw can be detected by the reduction of back wall reflection as well as the appearance of a feature before the first wall reflection. The fact that flaw A could be clearly detected by UT but not by ECT demonstrates that UT has significantly greater depth penetration compared with ECT. The results of this UT experiment on flaw A and B are consistent with the x-ray results in Fig. 5.

*D.    Surface inclusions*

Another type of flaws often found by ECT is the surface inclusions (Fig. 8(a)). Most of these inclusions have similar surface appearance as the wire, are difficult to distinguish by visual inspection. Our tensile fatigue tests showed that inclusions of > 0.5 mm in depth can cause significant reduction in wire's fatigue life. So it is necessary to inspect the wire for surface inclusions by NDT. ECT has been proven to be very effective. Fig. 8(b) shows an ECT trace of an inclusion. Both chemical analysis and magnetic imaging have determined that the inclusion is ferromagnetic. This explains the ability of finding very small inclusions by ECT.

From view point of long length wire inspection, ECT has the advantage of low cost, high speed and easy implementation, but with limited depth penetration ability. It is suitable for surface inspection and internal flaws of less than 3 mm deep. Meanwhile, UT has greater depth penetration but has issue of consistently coupling the transducer to the wire. Therefore further development in UT of the wires with large cross-section is needed.

By the time of writing this paper, we have inspected a few hundreds of meters of AL60 wires and identified a number of surface inclusions. Fig. 9 shows ECT surface inspection of an about 100 m long 13.7 mm diameter AL60 precursor rod. An encircling probe of 20.8 mm inner diameter 100 – 500 kHz was used. A plastic adaptor was made to fit the 13.7 mm diameter rod being inspected.

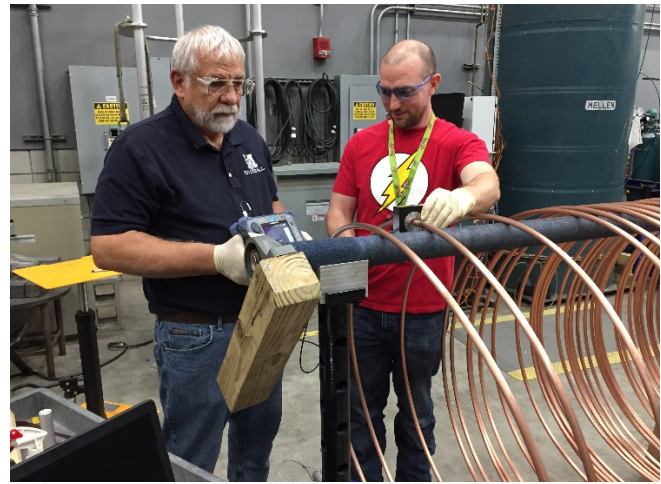

Fig. 9. ECT inspection of 13.7 mm diameter Glidcop AL60 precursor (about 100 m long). The encircling probe with frequency range of 100 -500 MHz was used.

IV.    CONCLUSION

We present experiments of NDT inspection of high strength conductor wires using ECT, UT and x-ray 2D and 3D methods. Internal chevron cracks were found in some Glidcop AL60 conductors by all three methods. Surface flaws were successfully detected by ECT method. The capability of long length ECT inspection of pulsed magnetic wires is developed.

V.    ACKNOWLEDGEMENT

We would like to thank Mr. Daniel Bone of Delphi Precision Imaging for performing x-ray 2D and 3D (CT) scans, Mr. William (Chuck) Edie of Olympus NDT for help on ECT of short samples, and Mr. Justin Deterding and Mr. Donald Richardson of the NHMFL for eddy current inspection of long length wires.